# Optical spectra, crystal-field parameters, and magnetic susceptibility of the new multiferroic NdFe$_3$(BO$_3$)$_4$

M. N. Popova,* E. P. Chukalina, and T. N. Stanislavchuk

*Institute of Spectroscopy, Russian Academy of Sciences, 142190 Troitsk, Moscow region, Russia*

B. Z. Malkin and A. R. Zakirov

*Kazan State University, 420008 Kazan, Russia*

E. Antic-Fidancev

*Laboratoire de Chimie Applique´e de l'E tat Solide, CNRS-UMR7574, ENSCP, 11, Rue Pierre et Marie Curie, F-75231 Paris Cedex 05, France*

E. A. Popova

*Physics Faculty, Moscow State University, 119992 Moscow, Russia*

L. N. Bezmaternykh and V. L. Temerov

*L.V. Kirensky Institute of Physics, Siberian Branch of RAS, Krasnoyarsk, 660036, Russia*

*popova@isan.troitsk.ru

**Abstract**

We report high-resolution optical absorption spectra for NdFe$_3$(BO$_3$)$_4$ trigonal single crystal which is known to exhibit a giant magnetoelectric effect below the temperature of magnetic ordering $T_N$ = 33 K. The analysis of the temperature-dependent polarized spectra reveals the energies and, in some cases, symmetries and exchange splittings of Nd$^{3+}$ 84 Kramers doublets. We perform crystal-field calculations starting from the exchange-charge model, obtain a set of six real crystal-field parameters, and calculate wave functions and magnetic $g$-factors. In particular, the values $g_\perp$= 2.385, $g_\parallel$= 1.376 were found for the Nd$^{3+}$ ground-state doublet. We obtain $B_{loc}$=7.88 T and |$J_{FN}$|= 0.48 K for the values of the local effective magnetic field at liquid helium temperatures at the Nd$^{3+}$ site and the Nd – Fe exchange integral, respectively, using the experimentally measured Nd$^{3+}$ ground-state splitting of 8.8 cm$^{-1}$. To check reliability of our set of crystal field parameters we model the magnetic susceptibility data from literature. A dimer containing two nearest-neighbor iron ions in the spiral chain is considered to partly account for quasi-one-dimensional properties of iron borates, and then the mean-field approximation is used. The results of calculations with the exchange parameters for Fe$^{3+}$ ions $J_{nn}$ = -6.25 K (intra-chain interactions) and $J_{nnn}$ = -1.92 K (inter-chain interactions) obtained from fitting agree well with the experimental data.

## I. INTRODUCTION

The borates RM$_3$(BO$_3$)$_4$ (R = Y or a rare earth, M = Al, Ga, Sc, Cr, Fe) have the structure without inversion symmetry, similar to that of the natural mineral huntite CaMg$_3$(CO$_3$)$_4$. Aluminum borates Nd$_x$(Y,Gd)$_{1-x}$Al$_3$(BO$_3$)$_4$ from this family, being used as nonlinear active media for self-frequency-doubling and self-frequency-summing lasers, were widely studied during decades (see, e.g., Refs [1-3]). The huntite-type iron borates have also been synthesized long ago [4] but only a recent improvement of crystal-growth technologies has opened possibilities for detailed studies of their physical properties. At present, the most comprehensively investigated iron borate crystal is GdFe$_3$(BO$_3$)$_4$. It was found that at 156 K GdFe$_3$(BO$_3$)$_4$ undergoes the structural phase transition [5] from the trigonal R32 structure into the less symmetric but also trigonal P3$_1$21 one [6], at $T_N$ =37 K it orders antiferromagnetically [5] into the collinear easy-plane structure [7], and at $T_R$ ≈10 K a reorientation of the magnetic moments into the easy-axis magnetic structure occurs [5, 7]. The structural phase transition is accompanied by the antiferroelectric ordering [6, 8] and, hence, GdFe$_3$(BO$_3$)$_4$ is a multiferroic (it possesses both magnetic and ferroelectric order parameters). Multiferroic features of GdFe$_3$(BO$_3$)$_4$ have nicely been shown in the recent studies of its magnetoelectric properties, by demonstrating a possibility to control the electric polarization by the magnetic field [9, 10]. Quite recent similar work on NdFe$_3$(BO$_3$)$_4$ has shown that this compound demonstrates appreciably greater than GdFe$_3$(BO$_3$)$_4$ magnetic-field-dependent electric polarization and a giant quadratic magnetoelectric efffect [11] and, thus, can be considered for device applications. The main difference between the two compounds is that Gd$^{3+}$ has zero orbital moment, L=0, in the ground state ($^8S_{7/2}$, separated by 32000 cm$^{-1}$ from the first excited state) and, consequently, in the first approximation is not influenced by the crystal field (CF), while the ground state of Nd$^{3+}$ in NdFe$_3$(BO$_3$)$_4$ ($^4I_{9/2}$, L=6, S=3/2, J=9/2) is split by CF into five Kramers doublets. The total CF splitting of the Nd$^{3+}$ ground state in oxides usually amounts to several hundreds of wave numbers (see, e.g., Ref [12]).

To account qualitatively for the crystal-field effects in NdFe$_3$(BO$_3$)$_4$, we have undertaken the broad-band polarized optical absorption study of the *f-f* transitions of the Nd$^{3+}$ ion in NdFe$_3$(BO$_3$)$_4$ and performed crystal-field calculations. The exchange-charge model was used to obtain the starting CF parameters, then a final fitting to the experimentally found energies of the CF levels was accomplished. To check the calculated magnetic g-factors for the CF levels of $Nd^{3+}$ in $NdFe_3(BO_3)_4$, we have analyzed temperature dependence of the magnetic susceptibilities measured recently [13] and estimated parameters of magnetic exchange interactions.

## II. EXPERIMENTAL DETAILS

$NdFe_3(BO_3)_4$ samples were grown using a $Bi_2Mo_3O_{12}$ – based flux, as described in Refs.[14, 15]. Unlike the $Bi_2O_3$ based fluxes, in $Bi_2Mo_3O_{12}$ - based flux $Bi_2O_3$ is strongly bonded to $MoO_3$ which excludes a partial substitution of bismuth for a rare earth during crystallization. Spontaneous nucleation from the flux resulted in small single crystals (1×1×1 mm). These crystals were used as seeds to grow large (10×5×5 mm) single crystals of $NdFe_3(BO_3)_4$. Crystals were dark-green in color, they had good optical quality. The samples were oriented using their morphology and optical polarization methods. For optical measurements, we used thin (0.15 mm) platelets cut either perpendicular or parallel to the c-axis of the crystal. High-resolution (up to 0.1 cm$^{-1}$) polarized temperature-dependent spectra were registered in the region 1500-25000 cm$^{-1}$ using BOMEM DA3.002 and Bruker 125HR high-resolution Fourier spectrometers. The sample was in a helium-vapor cryostat at a fixed (±0.15 K) temperature between 4.2 and 300 K. We used either unpolarized light propagating along the c-axis of the crystal (**k**||c, **E**, **H**⊥c — α-polarization) or linearly polarized light incident perpendicular to the c-axis (**k**⊥c, **E**||c — π-polarization, **E**⊥c — σ-polarization). $GdFe_3(BO_3)_4$ samples were also studied, to account for the broad absorption bands related to the $Fe^{3+}$ ions (the $Gd^{3+}$ ion has no optical transitions in the whole spectral range studied).

To facilitate the identification of the CF sublevels for several multiplets of $Nd^{3+}$, we compared spectra of $NdFe_3(BO_3)_4$ with those of $Nd_{0.1}Gd_{0.9}Al_3(BO_3)_4$. This latter crystal with the same structure as $NdFe_3(BO_3)_4$ does not order magnetically and, thus, its low-temperature spectra are not complicated by the line splittings due to exchange splittings of $Nd^{3+}$ Kramers doublets. Both $GdFe_3(BO_3)_4$ and $Nd_{0.1}Gd_{0.9}Al_3(BO_3)_4$ single crystals were grown in the same way as $NdFe_3(BO_3)_4$.

## III. STRUCTURE AND SYMMETRY CONSIDERATIONS

The crystal structure of $NdFe_3(BO_3)_4$ is described by the trigonal space group R32, down to at least 2 K [8]. The hexagonal unit cell contains three formula units. There is only one position for a rare-earth ion in this structure, its symmetry is $D_3$. The lattice structure constants were presented in Ref.[16]. Each of the first two coordination shells of the Nd site contains six oxygen ions at the distance $R_1$= 0.23858 nm and $R_2$= 0.31786 nm, respectively. These ions are arranged into deformed triangular prisms with the upper and lower triangles parallel to the crystallographic (001) plane and rotated relative to each other by 11.6 and 23.6 degrees in the first and the second coordination prisms, respectively. The third coordination shell contains six $Fe^{3+}$ ions at the distance R(NdFe) = 0.37846 nm which also form a deformed triangular prism with the rotation angle 40.3$^0$. The nearest $Nd^{3+}$ and $Fe^{3+}$ ions are bridged by the one of the nearest neighbor oxygen ions, the angle between the bonds $O^{2-}$ - $Nd^{3+}$ ($R_1$) and $O^{2-}$ - $Fe^{3+}$ ($R_3$ = 0.19797 nm) equals 2π/3. The crystal field of the $D_3$ symmetry splits multiplets (with a given value of a total momentum J) of a free $Nd^{3+}$ ion into J+1/2 Kramers doublets of the $\Gamma_4$ and $\Gamma_{56}$ symmetry (see the first column of Table 1). Fig. 1 displays the scheme of CF splittings and explains the line labeling in the experimental spectra presented below. Table 2 contains selection rules for optical transitions.

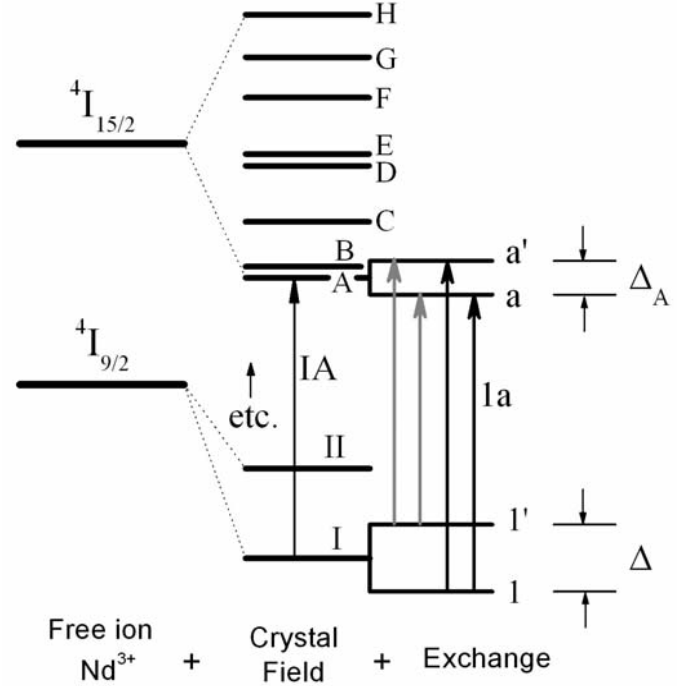

Fig.1. Scheme of $Nd^{3+}$ levels in a magnetic crystal and of optical transitions.

The nine iron sublattices form three chains twisted along the c-axis with the nearest-neighbor distance $R_{nn}$ = 0.31908 nm within a chain, the two next-nearest-neighbor iron ions are at the distance $R_{nnn}$ = 0.4403 nm in the neighbor chains. Each $Fe^{3+}$ ion has the two nearest $Nd^{3+}$ ions. The point symmetry at the iron sites is $C_2$. The $C_2$ axes are in the basis plane and rotate by 2π/3 along the chains. A fragment of the structure is shown in Fig. 2.

## IV. EXPERIMENTAL RESULTS

Narrow spectral lines due to the f-f optical transitions of $Nd^{3+}$ in $NdFe_3(BO_3)_4$ are superimposed onto broad absorption bands $^6A_1 \rightarrow {}^4T_1$ and $^6A_1 \rightarrow {}^4T_2$ in the near infrared and red spectral regions related to the d-d transitions of the $Fe^{3+}$ ions [17, 18] (see Fig. 3). We registered these bands in the spectra of $GdFe_3(BO_3)_4$ (the $Gd^{3+}$ ion has no energy levels up to about 33000 cm$^{-1}$) and subtracted them from the spectra of $NdFe_3(BO_3)_4$ to obtain the $Nd^{3+}$-related lines. Figs. 4-7 give a detailed picture of the spectra in different $Nd^{3+}$ multiplets.

As an example, Fig. 4 shows the α-polarized transmission spectra in several selected multiplets at different temperatures.

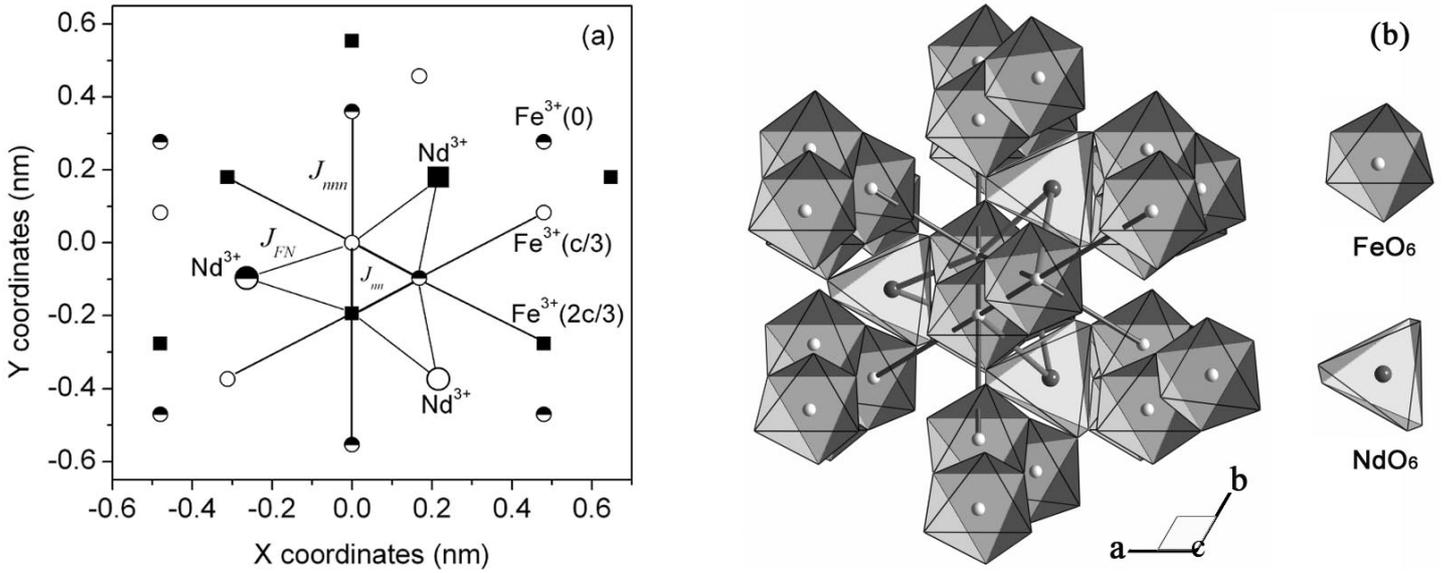

Fig.2. Projection of the magnetic ions of $NdFe_3(BO_3)_4$ onto the basis plane (z-coordinates are given in brackets) (a) and the corresponding three-dimensional fragment of the crystal structure (boron atoms are not shown) (b).

Absorption from the excited Stark levels of the ground multiplet $^4I_{9/2}$ results in spectral lines that disappear at low temperatures. They were used to find the Stark structure of the ground multiplet. Both the distances from the main lines and intensity dependences on the temperature were taken into account. Fig. 5 displays several more spectral multiplets in the paramagnetic phase of $NdFe_3(BO_3)_4$. Table 1 lists the positions of the CF energy levels in $NdFe_3(BO_3)_4$ found from the analysis of temperature-dependent polarized absorption spectra.

Below the temperature of the magnetic ordering $T_N$=33 K [17], the $Nd^{3+}$ Kramers doublets and, hence, the spectral lines split. Level splittings as determined from the spectra are also indicated in Table 1. We note that the positions of the two lowest energy levels in the $^4I_{15/2}$ CF multiplet of $NdFe_3(BO_3)_4$ could be found only from the low-temperature spectra, as centers of lines split in the magnetically ordered state. At 40 K, the transitions IA and IB superimpose and give only one line, while the spectra of the similar paramagnetic compound $Nd_{0.1}Gd_{0.9}Al_3(BO_3)_4$ clearly indicate the presence of the two lines (see Fig. 6 (b) and also Ref. [17]).

Now, where possible, we ascribe labels 4 or 56 to the energy levels (which show whether the wave functions transform according to the $\Gamma_4$ or, respectively, the $\Gamma_{56}$ irreducible representation (IRREP), making use of the polarized absorption spectra and selection rules of Table 2. As an example, spectra in all the three polarizations α, σ, and π are shown in Fig. 7 for the transitions from the ground $^4I_{9/2}$ multiplet to the $^4I_{11/2}$ and $^4I_{13/2}$ multiplets in paramagnetic $NdFe_3(BO_3)_4$ at the temperature 40 K (slightly above the temperature of the magnetic ordering). First of all, we note that if the ground state were $\Gamma_{56}$, several lines in each multiplet would disappear in α-polarization because $\Gamma_{56} \to \Gamma_{56}$ transitions are strictly forbidden in this polarization. It is not the case, and hence the ground state is $\Gamma_4$.

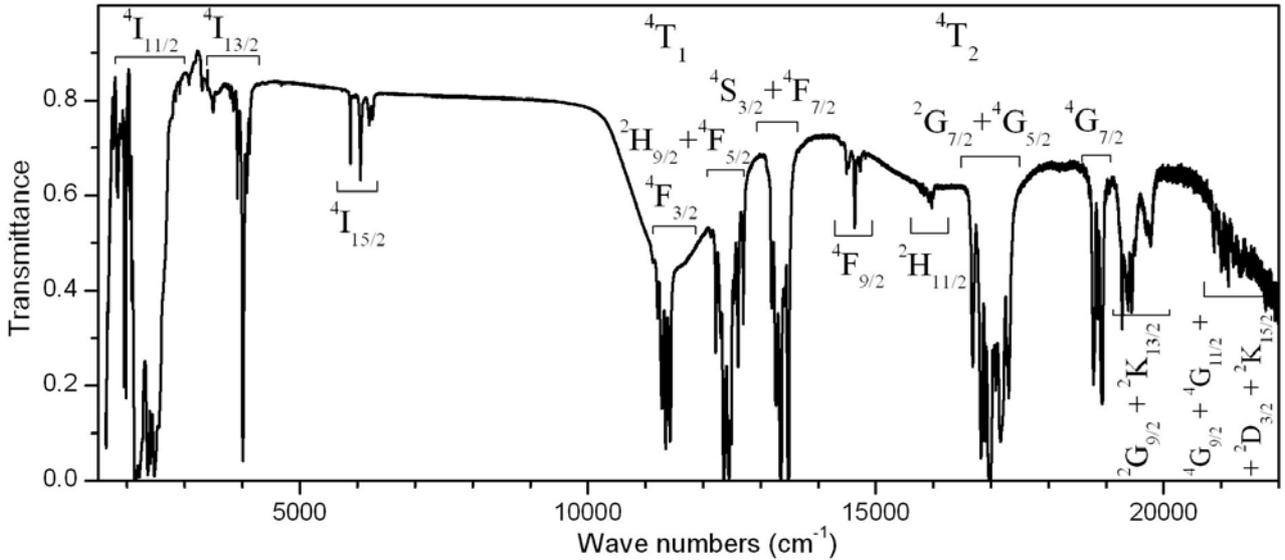

Fig.3. Transmission spectrum of paramagnetic $NdFe_3(BO_3)_4$ at $T$=40 K. The final states of the $Nd^{3+}$ ($Fe^{3+}$) transitions from the ground state $^4I_{9/2}$ ($^6A_1$) are indicated.

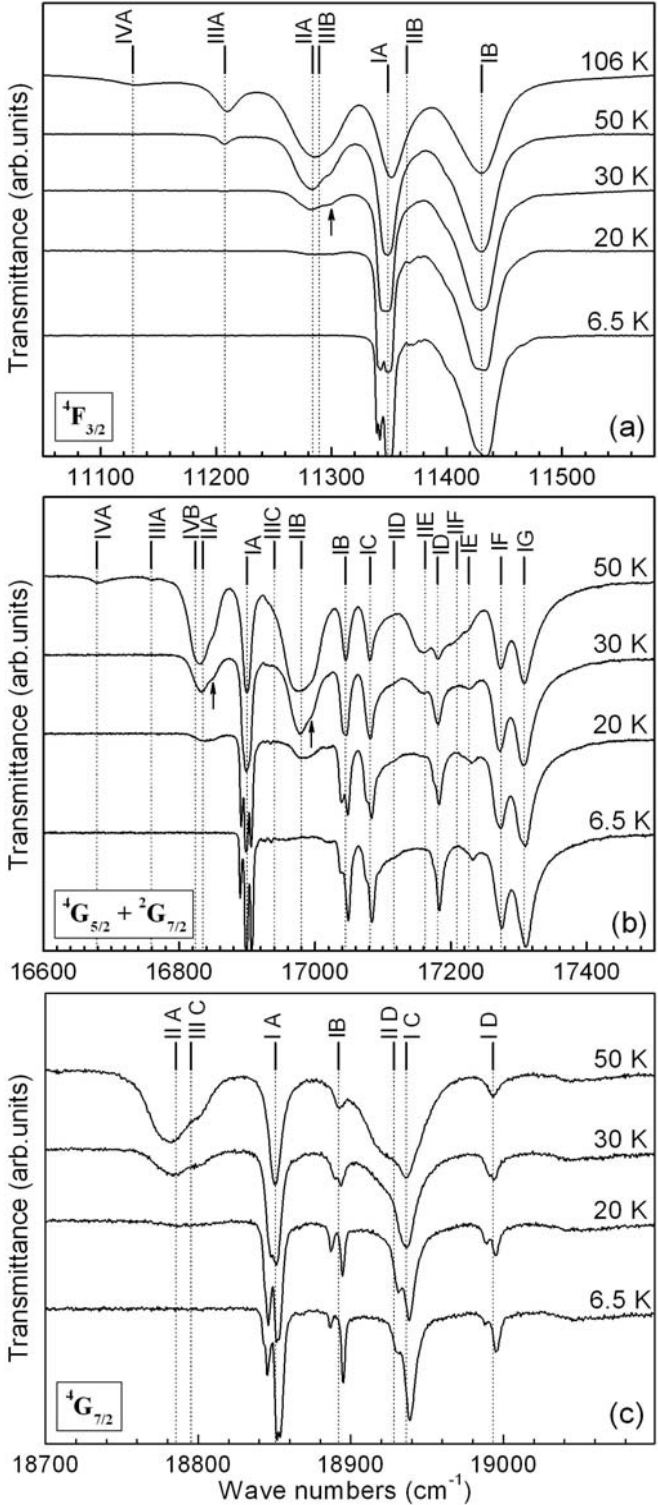
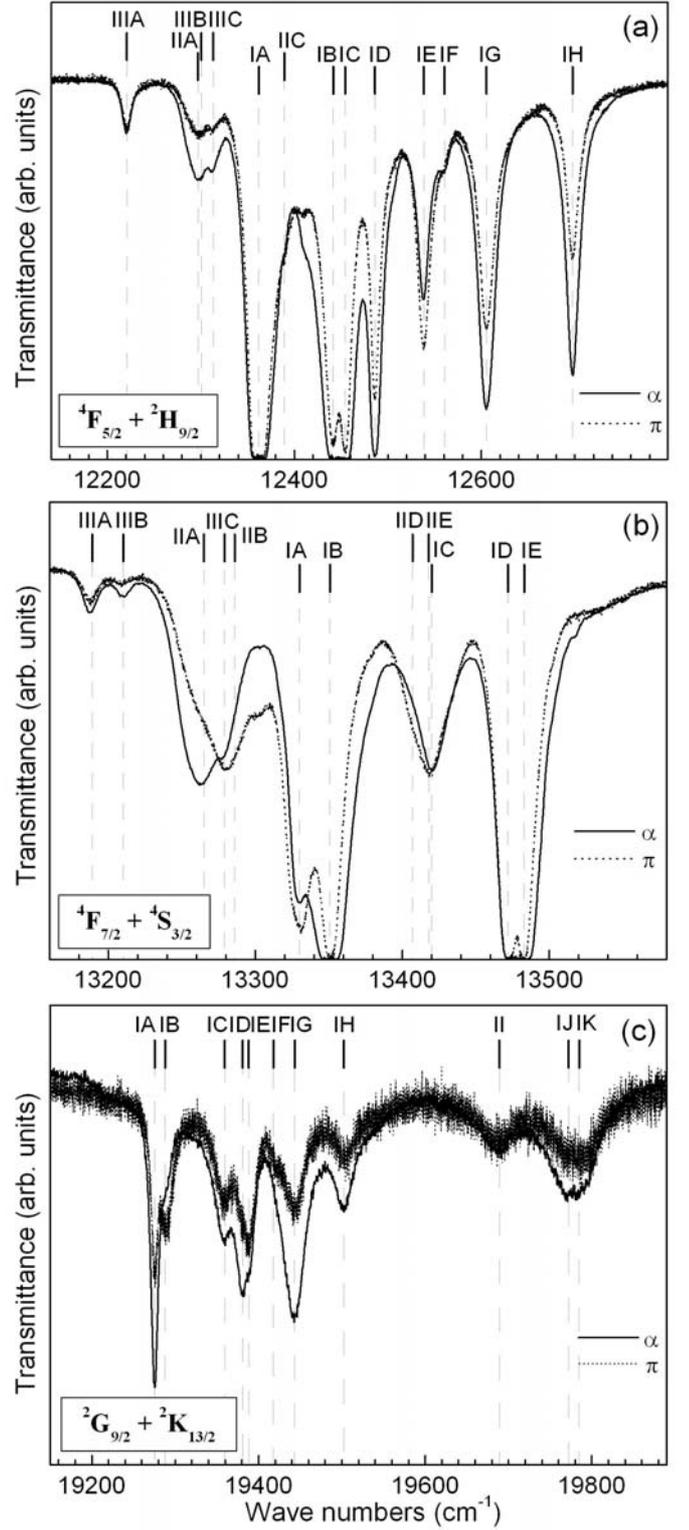

Fig.4. Transmission spectra of NdFe$_3$(BO$_3$)$_4$ at different temperatures for transitions in Nd$^{3+}$ from the ground state $^4I_{9/2}$ to different excited states. Identification of spectral transitions is given according to the scheme of Fig.1.

Besides, the assumption of the $\Gamma_{56}$ ground state does not allow conforming with both the selection rules of Table 2 and the first column of Table 1. On the contrary, certain transitions from the first excited state vanish just in α-polarization, indicating its $\Gamma_{56}$ symmetry.

Fig.5. Transmission spectra of paramagnetic NdFe$_3$(BO$_3$)$_4$ at T= 40 K in α and π polarizations.

The lowest in energy $^4I_{9/2} \rightarrow {}^4I_{11/2}$ transition allowed in the free Nd$^{3+}$ ion as the magnetic dipole one, contains a considerable magnetic dipole contribution. In particular, the IB line is purely magnetic dipole which follows from the fact that π- and α-polarized spectra coincide (Fig. 7 (a)). For most of the higher transitions, α- and σ-polarized spectra are very close, that is electric dipole transitions dominate. However, even in this case it is not possible to assign IRREPs from the σ-polarized main line (see Table 2). Here, the low-frequency satellite displaced by 65 cm$^{-1}$ (energy of the first excited state having the $\Gamma_{56}$ symmetry) from the main line can help. Namely, its σ- (π-) polarization points to the $\Gamma_4$ ($\Gamma_{56}$) symmetry for the terminal level of the transition. Unfortunately, (i) not always such satellites are observed and (ii) magnetic dipole contributions are present even for other than $^4I_{9/2} \to {^4I_{11/2}}$ transitions. That is why not all the IRREPs of Table 1 could be assigned unambiguously from the experiment.

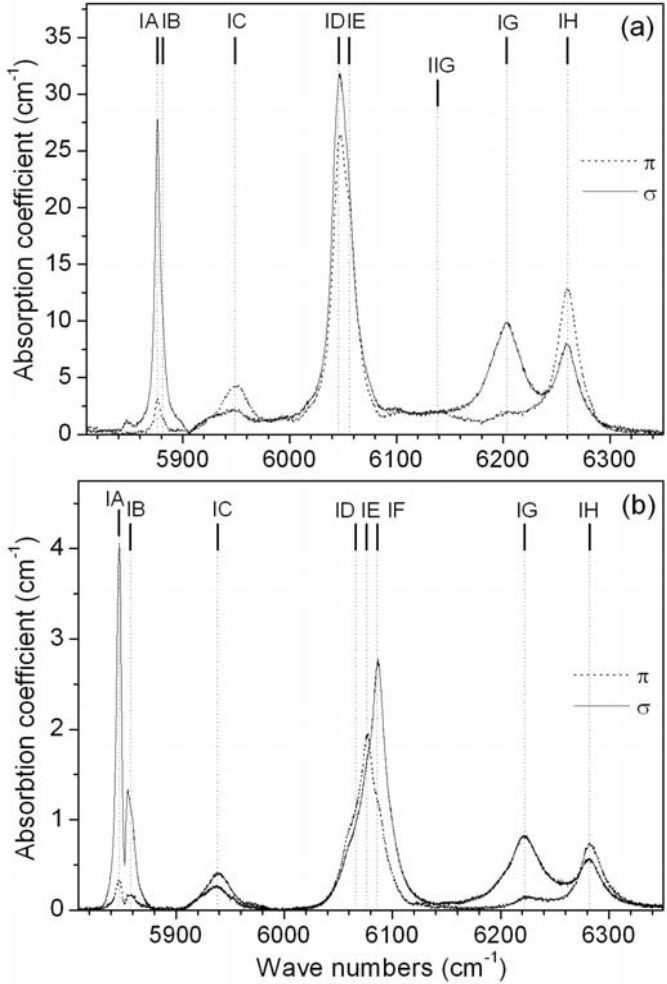

Fig.6. Comparison of the $Nd^{3+}$ $^4I_{9/2} \to {^4I_{15/2}}$ optical transitions in paramagnetic $NdFe_3(BO_3)_4$ and $Nd_{0.1}Gd_{0.9}Al_3(BO_3)_4$.

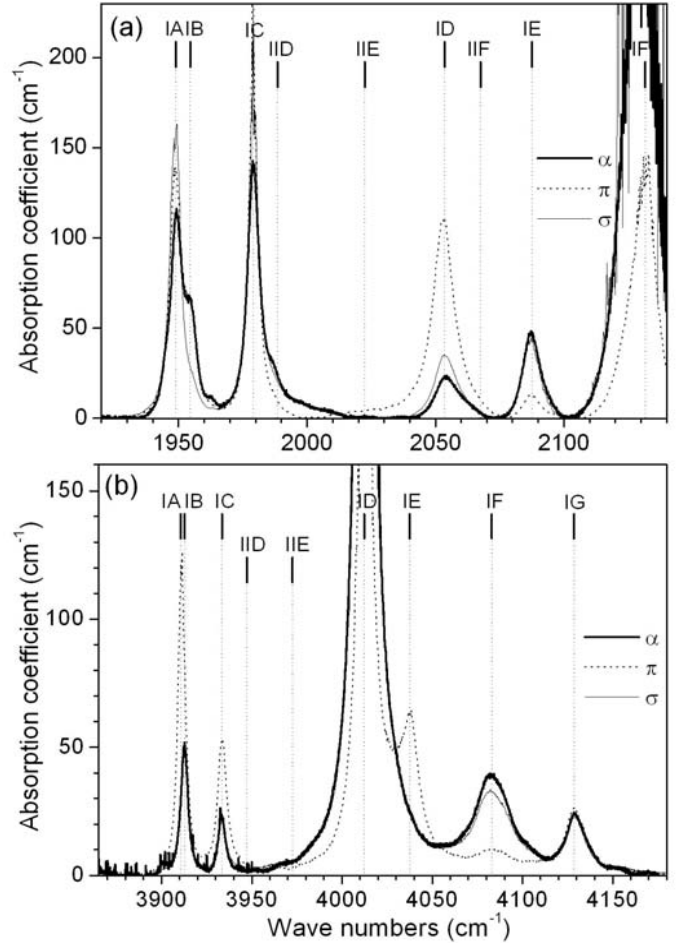

Fig.7. Absorption in $Nd^{3+}$ $^4I_{9/2} \to {^4I_{11/2}}$ (a) and $^4I_{9/2} \to {^4I_{13/2}}$ (b) transitions in paramagnetic $NdFe_3(BO_3)_4$ in all the three polarizations.

Our high-resolution spectra deliver information on line width and line shapes. The lowest levels in each multiplet are the most narrow, while the highest states suffer from additional one-phonon relaxation. Components of the split low-frequency lines at the temperatures lower than 5 K are as narrow as 1-2 cm$^{-1}$. This temperature-independent width is the inhomogeneous width due to lattice defects. In the paramagnetic state, disordered iron magnetic moments impose a fluctuating magnetic field at the $Nd^{3+}$ sites that causes a fluctuating Zeeman splitting of $Nd^{3+}$ Kramers doublets and, hence, a line broadening. A noticeable additional broadening is observed for the lines corresponding to the transition from the ground state to the levels lying higher than $\upsilon_0 = 65$ cm$^{-1}$ from the lowest level in a multiplet. The value of $\upsilon_0$ correlates with the lowest phonon frequency observed in the infrared reflection spectra of $NdFe_3(BO_3)_4$ [19]. It is interesting to note that in $NdFe_3(BO_3)_4$ transitions from the first excited state (65 cm$^{-1}$) are broader than those from the next excited states (141 and 221 cm$^{-1}$) and have a doublet structure (see, e.g., Figs 4 (a, b)). We suppose that this is caused by the resonance interaction between the $Nd^{3+}$ CF level and lattice vibrations.

## V. CRYSTAL FIELD CALCULATIONS

The measured energy spectrum of $Nd^{3+}$ ions in $NdFe_3(BO_3)_4$ was analyzed using the Hamiltonian

$$H = H_{FI} + H_{CF} \qquad (1)$$

where $H_{FI}$ is the effective free-ion Hamiltonian, and $H_{CF}$ corresponds to the interaction of the $Nd^{3+}$ ground $4f^3$ configuration with the crystal field. The electrostatic interaction between $4f$ electrons determined by Racah parameters $E_1$, $E_2$, $E_3$, spin-orbit interaction (with the coupling constant ξ), Trees' two-particle (α, β, γ parameters) and Judd' three-particle (parameters $T^2$-$T^4$, $T^6$-$T^8$) corrections to the electrostatic interaction were taken into account in the free-ion Hamiltonian. At the $Nd^{3+}$ sites with the $D_3$ symmetry, in the Cartesian system of coordinates with the z-axis along the crystallographic *c*-axis and the *x*-axis along the crystallographic *a*-axis (which is the $C_2$ symmetry axis (see Fig. 2), the crystal field can be defined by six independent real CF parameters $B_q^p$: ($p = 2,4,6$; $p \geq q = 0,-3,6$): (2)

$$H_{CF} = \sum_i [B_0^2 C_0^{(2)}(i) + B_0^4 C_0^{(4)}(i) + B_{-3}^4 C_{-3}^{(4)}(i) + B_0^6 C_0^{(6)}(i) + B_{-3}^6 C_{-3}^{(6)}(i) + B_6^6 C_6^{(6)}(i)].$$

Here the sum is taken over 4$f$ electrons, $C_q^{(p)}(i)$ is the spherical tensor operator of rank $p$. The absolute value of the parameter $B_0^2 \sim 550$ cm$^{-1}$ is obtained directly from the measured splitting of the $^4F_{3/2}$ multiplet (see Table 1), the positive sign of this parameter and initial estimations of parameters $B_q^4, B_q^6$ were found from calculations in the framework of the exchange-charge model following the procedure used earlier in the analysis of the Nd$^{3+}$ CF spectra in Nd$_2$BaNiO$_5$ [20]. In such a way we obtained $B_0^4 = -568 - 107 \cdot G$, $B_{-3}^4 = 317 + 41 \cdot G$, $B_0^6 = 67 + 70 \cdot G$, $B_{-3}^6 = 1.4 + 16 \cdot G$, $B_6^6 = 64 + 38 \cdot G$ (cm$^{-1}$), where the first and second terms correspond respectively to contributions from the effective point ion charges (-1 for oxygen ions and +1.5 for each metal ion, in units of proton charge) and from the exchange charges proportional to the single phenomenological model parameter $G_\sigma = G_\pi = G$. The value of the parameter $G = 7$ was found from a comparison of the calculated and measured CF splittings of the ground multiplet. The final set of parameters of the Hamiltonian (1), operating in the total basis of 364 states of the electronic 4$f^3$ configuration, $E_1 = 4775.5$, $E_2 = 23.21$, $E_3 = 484.72$, $\alpha = 21.5$, $\beta = -626$, $\gamma = 1500$, $\xi = 873.7$, $T_2 = 250$, $T_3 = 40$, $T_4 = 80$, $T_6 = -274$, $T_7 = 316$, $T_8 = 271$, $B_0^2 = 551$, $B_0^4 = -1239$ (- 1317), $B_{-3}^4 = 697$ (604), $B_0^6 = 519$ (557), $B_{-3}^6 = 105$ (113), $B_6^6 = 339$ (330) cm$^{-1}$, was obtained by fitting the eigenvalues of (1) to the measured CF energies (the initial values of CF parameters are in brackets). The measured CF energies of the $^2H_{11/2}$ multiplet were not taken into account in the fitting procedure. Similarly to Ref. [21], we have calculated the Stark structure of this multiplet using four times diminished reduced matrix elements of the $C_q^{(4)}$ operators. The calculated CF energies are compared with the experimental data in Table 1. The same Table shows the IRREPs corresponding to the calculated wave functions and how do they correlate with the experimental data.

The magnetic spectroscopic factors $g_{\alpha\beta} = \delta_{\alpha\beta} g_{\alpha\alpha}$; $g_{xx} = g_{yy} = g_\perp$, $g_{zz} = g_\parallel$ were computed for each doublet CF state $j$ using the eigenfunctions $|j\pm\rangle$ of the Hamiltonian (1):

$$g_{\alpha\alpha} = 2(|\langle j+|\sum_i l_\alpha(i) + 2s_\alpha(i) | j-\rangle|^2 + |\langle j+|\sum_i l_\alpha(i) + 2s_\alpha(i) | j+\rangle|^2)^{1/2}.$$ (3)

Here $l$ and $s$ are the one-electron orbital and spin moments, respectively. Results of calculations are presented in Table 1. It is worth noting that for $\Gamma_{56}$ levels $g_\perp = 0$ by symmetry considerations.

Table 1 shows that the calculated splittings of the Nd$^{3+}$ multiplets with energies below 15000 cm$^{-1}$ (positions of the CF states with higher energies are more dependent on the mixing with the excited configurations) and symmetry properties of the CF sublevels in the paramagnetic phase of NdFe$_3$(BO$_3$)$_4$ agree satisfactorily with the experimental data. However, relative positions and energies of different sublevels do not change critically when varying the CF parameters up to 10 %. To additionally check reliability of the obtained set of the CF parameters, in the next Section we analyze susceptibility data presented recently in Ref. [13].

## VI. MAGNETIC SUSCEPTIBILITY OF NdFe$_3$(BO$_3$)$_4$

First of all, it is necessary to clear up the question whether the essential anisotropy of the calculated *g*-tensor in the ground multiplet of the Nd$^{3+}$ ions is consistent with a very weak anisotropy of the measured susceptibility tensor in the paramagnetic phase. The experimental data on the transversal and longitudinal magnetic susceptibilities of NdFe$_3$(BO$_3$)$_4$ [13] are presented in Fig 8 (symbols). The Nd$^{3+}$ single-ion susceptibility tensor is diagonal in the system of coordinates which has been used in calculations of the CF parameters. The diagonal components, $\chi_{xx}^{Nd} = \chi_{yy}^{Nd}$ and $\chi_{zz}^{Nd}$, were computed for different temperatures using the CF parameters presented above. As it is seen in Fig. 8, the calculated differences $\Delta\chi(T) = N_A[\chi_{xx}^{Nd}(T) - \chi_{zz}^{Nd}(T)]$ (where $N_A$ is the Avogadro number), are close to the measured differences between $\chi_\perp$ and $\chi_\parallel$ in the paramagnetic phase. However, the calculated Nd$^{3+}$ single ion susceptibilities grow much faster than the experimental ones when the temperature is lowered. Evidently, the susceptibilities of the neodymium and iron subsystems are renormalized due to magnetic interactions between Nd$^{3+}$ and Fe$^{3+}$ ions.

To describe the measured susceptibilities, we shall consider the following Hamiltonian of the exchange interactions between the nearest- and the next-nearest-neighbor Fe$^{3+}$ ions and between the nearest Fe$^{3+}$ and Nd$^{3+}$ ions:

$$H_{exch} = -2J_{nn} \sum_{Ck} S_{k,C} S_{k+1,C} - 2J_{nnn} \sum_{C<C',kp} S_{k,C} S_{p,C'} - 2J_{FN} \sum_{Ckp} S_{k,C} S_{Nd,p}.$$ (4)

Here, $S_{k,C}$ ($S=5/2$) is the spin moment of a Fe$^{3+}$ ion in the chain $C$, $S_{Nd,p}$ is the total spin moment of the Nd$^{3+}$ ion ($S$(Nd$^{3+}$)=($g_J$-1)$J$ in the space of states of a multiplet with the total angular momentum $J$ and the Lande factor $g_J$), $J_{nn}$, $J_{nnn}$, and $J_{FN}$ are the Fe-Fe nearest-neighbor, Fe-Fe next-nearest-neighbor, and Nd-Fe nearest-neighbor exchange interaction parameters, respectively, the corresponding restrictions are implied in the sums. The nearest Fe-Fe neighbors are inside one iron chain, while the next-nearest ones are in neighboring chains, and their z-coordinates differ by $c/3$. In Eq. (4), positive (negative) signs of exchange parameters correspond to the ferromagnetic (antiferromagnetic) exchange interaction. At temperatures below 400 K, only the Nd$^{3+}$ ground multiplet $^4I_{9/2}$ contributes to the susceptibility. Of course, restriction by isotropic exchange interactions is a very crude approximation. In the magnetically ordered state, magnetic moments of Fe$^{3+}$ ions lie in the planes normal to the *c*-axis [16]. As it follows from a comparison between the calculated *g*-factors and the measured splittings Δ of the CF doublet sublevels of the excited multiplets $^4I_{13/2}$ and $^4I_{15/2}$ in the internal magnetic fields (see Table 1), the effective local magnetic fields at the neodymium sites are different for different sublevels and may have both transversal and longitudinal (along the *c*-axis) components.

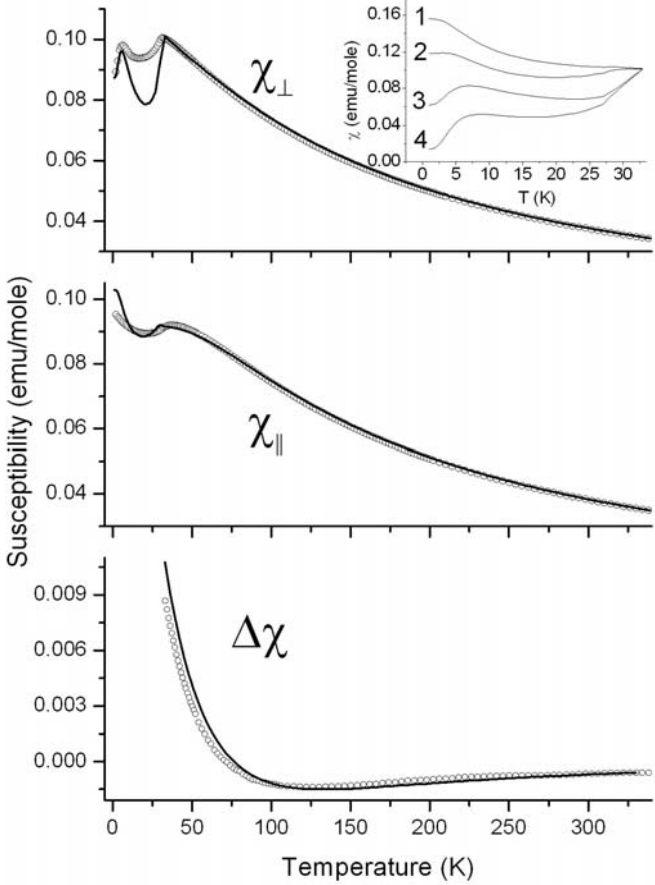

Fig.8. Temperature dependences of magnetic susceptibilities of NdFe$_3$(BO$_3$)$_4$ (symbols correspond to experimental data [13], solid curves present results of calculations). Inset displays the calculated susceptibilities of domains with the spontaneous magnetic moments in the crystallographic basis plane declined by the angle $\phi$ from the external magnetic field (1 - $\phi=\pi/2$, 2 - $\phi=\pm\pi/3$, 3 - $\phi=\pm\pi/6$, 4 - $\phi=0$).

This means that the generalized Hamiltonian corresponding to the exchange interaction between Fe$^{3+}$ and Nd$^{3+}$ ions contains not only bilinear terms in spin operators $S$(Fe$^{3+}$) and $S$(Nd$^{3+}$) but also direct products of $S$(Fe$^{3+}$) and spherical tensor operators $\sum_i C^{(p)}(i)$ of ranks $p=3, 5$ (the label *i* numerates 4*f* electrons localized at the Nd$^{3+}$ ion). However, the relative role of the anisotropic terms is essentially diminished in the case of the ground Nd$^{3+}$ multiplet due to higher value of the factor (1-$g_J$) and lower values of matrix elements for operators $\sum_i C^{(p)}(i)$. For instance, the calculated maximum splitting of a doublet state due to the perturbation $\sum_i C^3_{-3}(i)$ is 11 times less in the $^4I_{9/2}$ multiplet than in the $^4I_{13/2}$ one.

The absolute value of the exchange integral $J_{FN}$ can be obtained from the splitting of the Nd$^{3+}$ ground doublet $\Delta_0 = g_\perp \mu_B B_{loc}$ =8.8 cm$^{-1}$ ($\mu_B$ is the Bohr magneton) measured here and in Ref. [17] at 4.2-5 K. In the magnetically ordered state with the average Fe$^{3+}$ ion spin moment $<S>$, the local magnetic field equals $B_{loc}$=12(1-$g_J$)|$J_{FN}$|$<S>/g_J\mu_B$. This field corresponds to the magnetic structure with the propagation vector $\mathbf{k}$ = [0 0 3/2] determined in Ref. [16], the magnetic moments of all the Fe$^{3+}$ and the Nd$^{3+}$ ions being parallel to the basis plane and having opposite directions in the planes (00z) and (0 0 z±1/3).Using the measured value $<S>$=4.9/2 [16] of the Fe$^{3+}$ spin moment, the value $g_J$=8/11 for the Nd$^{3+}$ ground state, and the calculated value of $g_\perp$ (see Table 1), we obtain $B_{loc}$=7.88 T and |$J_{FN}$|= 0.48 K. (Below, we shall also determine the sign of the exchange parameter $J_{FN}$ ). It is worth mentioning that the obtained value of the local magnetic field at the Nd$^{3+}$ site in the magnetically ordered state of NdFe$_3$(BO$_3$)$_4$ at 5 K is equal to the external magnetic field $B_a$ that saturates the longitudinal electric polarization in the magnetoelectric experiments of Ref. [9].

As the first step, we study the contribution of the iron subsystem into the magnetic susceptibility. The most simple approach is to consider a single iron ion and then to use the mean-field approximation (MFA). A total splitting of the iron multiplet $^6S_{5/2}$ in the crystal field usually is much less than one wave number. Neglecting the fine structure of this multiplet, we obtain the single iron ion isotropic magnetic susceptibility $\chi_0^{Fe} = \frac{S(S+1)}{3kT}(g\mu_B)^2$, where $g$=2 is the spin *g*-factor and *k* is the Boltzmann constant. In the framework of the MFA, the effective single iron ion Hamiltonian in the external magnetic field $\mathbf{B}$ can be written as $H^s = -g\mu_B \mathbf{SB} - 2J_s <\mathbf{S}> \mathbf{S}$, where $J_s$=2$J_{nn}$+2$J_{nnn}$. The corresponding effective susceptibility equals

$$\chi_s^{Fe} = \frac{\chi_0^{Fe}}{1 - 2\chi_0^{Fe} J_s /(g\mu_B)^2} \quad (5)$$

Again, as it was in the case of the Nd$^{3+}$ single ion magnetic susceptibility, $\chi_s^{Fe}$ grows with decreasing the temperature much more rapidly than the measured susceptibility.

This approximation is valid at temperatures much higher than $T_N$, but it fails close to $T_N$ due to strong fluctuations of iron magnetic moments in the chains. We can, at least partly, account for the specific properties of the quasi-one-dimensional system by considering the exact form of the susceptibility of a dimer containing two nearest-neighbor iron ions in the chain and only then applying the MFA. The eigenvalues of the Hamiltonian of a dimer in the external magnetic field $\mathbf{B}\|z$,

$$H^d = -2J_{nn}\mathbf{S_1S_2} - [g\mu_B B + 2(J_{nn} + 2J_{nnn})<S_z>](S_{1z}+S_{2z}) \quad (6)$$

Equal

$$\varepsilon(S_t, S_{tz}) = -S_t(S_t+1)J_{nn} - [2(J_{nn}+2J_{nnn})<S_z>+g\mu B]S_{tz},$$
$$(S_t=0,1,\ldots 2S), \quad (7)$$

where $<S_z>=<S_{1z}+S_{2z}>/2$ = Sp[$S_{tz}$exp(-$H^d/kT$)]/2Sp[exp(-$H^d/kT$)] is the average spin moment of an ion. We note that here the exchange parameter $J_{nn}+2J_{nnn}$ is responsible for an effective interaction of the dimer with its neighbors. Now it is easy to obtain the magnetic susceptibility per iron ion in the dimer ($g\mu_B<S_z>=\chi_d^{Fe}B$):

$$\chi_d^{Fe} = \frac{\chi_d}{1-2\chi_d(J_{nn}+2J_{nnn})/(g\mu_B)^2} \quad (8)$$

Here

$$\chi_d(T) = \frac{(g\mu_B)^2}{6kTZ(T)} \sum_{S_t=0}^{2S} S_t(S_t+1)(2S_t+1)\exp[J_{nn}S_t(S_t+1)/kT] \quad (9)$$

$$Z(T) = \sum_{S_t=0}^{2S} (2S_t+1)\exp[J_{nn}S_t(S_t+1)/kT] \quad (10)$$

To summarize, Eq. (9) presents the exact solution for the dimer. Applying the MFA to the iron dimer (instead of the single iron ion) results in Eq. (8) (to compare with Eq. (5) for the single ion in the effective field). We shall show later that Eq. (8), in contrast to Eq. (5), gives possibility to describe the experimentally observed temperature dependence of the magnetic susceptibility of $NdFe_3(BO_3)_4$. It should be noted that we have also analyzed the susceptibility of the four-particle cluster, some results will be given later.

To make a comparison with the experimental data, it is necessary to add the contribution from the neodymium subsystem and to take into account the interaction between the iron and the neodymium magnetic subsystems. We write the MFA free energy per unit cell (with the volume $v$) of the spherical paramagnetic crystal in the magnetic field $B$ in the form

$$F = \sum_{j\alpha\beta} m_{j\alpha}(2\chi_j)^{-1}_{\alpha\beta}m_{j\beta} - B\sum_j m_j -$$
$$2\sum_{jj'}\frac{(g_J-1)J_{FN}^{jj'}}{gg_J\mu_B^2}m_{jFe}m_{j'Nd} - \quad (11)$$
$$\frac{1}{2}\sum_{jj'\alpha\beta}m_{j\alpha}(Q_{j\alpha,j'\beta}-\frac{4\pi}{3v}\delta_{\alpha\beta})m_{j'\beta}.$$

Here $m_j$ is the magnetic moment of an ion in the $j$-sublattice (thus, both $Fe^{3+}$ and $Nd^{3+}$ ions are included), $\chi_j$ is the effective single ion susceptibility. The third term describes the Fe – Nd interactions. The last term corresponds to the magnetic dipole-dipole interactions, $Q_{j\alpha j'\beta}$ are the lattice sums (parameters of the Lorentz field) which have been computed by the Ewald method. From the minimum conditions $\frac{\partial F}{\partial m_{j\alpha}}=0$, we obtain the system of self-consistent linear equations relative the magnetic moments. A solution of these equations gives the susceptibility per mole as

$$\chi_{\alpha\beta}(T) = \frac{1}{3}N_A\sum_{jj'}K^{-1}_{j\alpha\,j'\beta}\chi_{\beta\beta}^{j'}(T) \quad (12)$$

where

$$K_{j\alpha\,j'\beta} = \delta_{jj'}\delta_{\alpha\beta} - \chi_{\alpha\alpha}^j(T)[Q_{j\alpha\,j'\beta}-(\frac{4\pi}{3v}+\frac{2(1-g_J)J_{FN}^{jj'}}{gg_J\mu_B^2})\delta_{\alpha\beta}] \quad (13)$$

In particular, when neglecting dipole-dipole interactions, we obtain

$$\chi_{\alpha\alpha}(T) = N_A[1-12\chi_d^{Fe}(T)\chi_{\alpha\alpha}^{Nd}(T)\left(\frac{2(1-g_J)J_{FN}}{gg_J\mu_B^2}\right)^2]^{-1} \times$$
$$[3\chi_d^{Fe}(T)+\chi_{\alpha\alpha}^{Nd}(T)-12\chi_d^{Fe}(T)\chi_{\alpha\alpha}^{Nd}(T)\frac{2(1-g_J)J_{FN}}{gg_J\mu_B^2}]. \quad (14)$$

Values of the exchange parameters $J_{nn}$ = -6.25 K, $J_{nnn}$ = -1.92 K, and the sign of the Nd-Fe exchange parameter $J_{FN}$ = +0.48 K were obtained from fitting the calculated temperature dependence of the transversal paramagnetic susceptibility $\chi_\perp(T)$ to the measured data. The obtained exchange parameters $J_{nn}$ and $J_{nnn}$ correspond to the antiferromagnetic exchange interactions between the $Fe^{3+}$ ions. The positive sign of the parameter $J_{FN}$ points to the ferromagnetic $Fe^{3+}$ - $Nd^{3+}$ exchange, as far as the spin moments are concerned. Because of the already mentioned relation $S(Nd^{3+})=(g_J-1)J$, the total angular momentum $J$ of the $Nd^{3+}$ ion in the ground state ($g_J$ = 8/11) is directed opposite to the $Nd^{3+}$ spin moment. We hence conclude that the exchange interaction between the $Fe^{3+}$ and the $Nd^{3+}$ ions tends to arrange their magnetic moments antiferromagnetically. Using the same exchange parameters, we have further calculated the longitudinal paramagnetic susceptibility $\chi_\parallel(T)$. As one can see from Fig.8, results of calculations agree well with experimental data in the paramagnetic phase ($T$ > 33 K).

We have thus shown that the model of a dimer (consisting of two neighboring iron ions along the chain) interacting with neighboring neodymium ions allows to describe the experimentally measured magnetic susceptibility of $NdFe_3(BO_3)_4$ in the paramagnetic region. The anisotropy of the magnetic susceptibility of the $Nd^{3+}$ ions in the crystal field of $NdFe_3(BO_3)_4$ is slightly reduced by the Nd-Fe interaction.

As the next step, we have considered the magnetically ordered phase. At temperatures below $T_N$, the susceptibilities $\chi_\perp$ and $\chi_\parallel$ of $NdFe_3(BO_3)_4$ in the external magnetic field parallel to the $a$ and $c$ axes, respectively, were calculated in the framework of the same dimer model with taking into account the spontaneous ordering of $Fe^{3+}$ spin moments. In this case the Hamiltonian (6) of a dimer contains additional terms:

$$H^d = -2J_{nn}S_1S_2 - [2(J_{nn}+2J_{nnn})<S>_0 +$$
$$4J_{FN}<S_{Nd}>_0](S_1-S_2) - \quad (15)$$
$$[g\mu_B B+2(J_{nn}+2J_{nnn})<\delta S>+4J_{FN}<\delta S_{Nd}>](S_1+S_2).$$

Here $<\delta S>$ and $<\delta S_{Nd}>$ are the average spin moments of $Fe^{3+}$ and $Nd^{3+}$ ions, respectively, induced by the external field $B$, $<S>_0$ is the spontaneous spin moment of $Fe^{3+}$ ions. The absolute values of $<S>_0$ at temperatures $T<T_N$ were determined in Ref. [16]. We used these values in calculations of $Nd^{3+}$ spin moments $<S_{Nd}>_0$ induced by the internal exchange field $B_i = 12(g_J-1)J_{FN}<S>_0/g_J\mu_B$ at the same temperatures. Assuming the multidomain magnetic structure with a statistical domain distribution, we considered three types of domains with the directions of spontaneous moments $<S>_0$ along three equivalent $C_2$ axes in the basal plane. For each domain, susceptibilities per $Fe^{3+}$ ion in a dimer and of $Nd^{3+}$ ion in the internal field were calculated, and the total susceptibility was obtained from Eq. (14). Results of calculations are presented in Fig.8 (we note that for the external magnetic field along the $c$-axis, all domains are equivalent). The obtained averaged susceptibilities qualitatively agree with the measured data at temperatures between 2 K and $T_N$. The position and shape of the low-temperature maximum in the transversal susceptibility $\chi_\perp$ (due to the splitting of the $Nd^{3+}$ ground doublet in the internal magnetic field) is well reproduced by calculations.

However, as one can see from Fig. 8, there are noticeable differences between the calculated and measured values of the transversal susceptibility at temperatures close to 20 K and of the longitudinal susceptibility at temperatures below 10 K. We obtained the same results in the framework of a more elaborated model considering tetramers (clusters which contain four adjacent $Fe^{3+}$ ions) in the iron chains and using slightly diminished exchange parameters ($J_{nn}$ = -6.07 K, $J_{nnn}$ = -1.65 K) to describe the temperature dependences of susceptibilities in the paramagnetic phase. The most important reason for a discrepancy between the measured and calculated susceptibilities must be connected with the fact that the MFA fails at temperatures close to $T_N$. To be specific, using the intra-chain and inter-chain exchange integrals $J_{nn}$ and $J_{nnn}$ given above, and the condition that the response of the iron subsystem to the staggered magnetic field (the antiferromagnetic susceptibility $\chi_{Fe}^{AF}(T)$) diverges at the transition temperature, we have obtained essentially overestimated values of $T_N$ = 82 K, 77 K, and 67.4 K for the two-, three-, and four-particle cluster models, respectively (the analytical forms of the antiferromagnetic susceptibilities for two- and three-particle clusters are presented in the Appendix). This result is a consequence of the non-zero transition temperatures for a single chain predicted by the considered cluster models (55 K, 46 K, and 39 K, respectively), although the one-dimensional Heisenberg chain should have no long-range order at any temperature. So, to clear up reasons for the remaining differences between the calculated and measured susceptibilities in the magnetically ordered phase, it is necessary to perform more detailed experimental and theoretical studies of magnetic properties of the iron subsystem.

## VII. CONCLUSIONS

We have found the energies of $Nd^{3+}$ CF levels (Kramers doublets) in the iron borate $NdFe_3(BO_3)_4$ analyzing our high-resolution temperature-dependent polarized absorption spectra. When possible, we ascribed the irreducible representations ($\Gamma_4$ or $\Gamma_{56}$ of the local point symmetry group $D_3$) to these levels and also determined their splitting in the magnetically ordered state. These data were used to carry out the crystal field calculations. The initial set of CF parameters obtained in the framework of the exchange charge model was then refined in a fitting procedure to the experimentally measured energy levels. Such approach permitted to avoid getting into a local minimum during the fitting procedure and to obtain physically reasonable and reliable set of the CF parameters. We have calculated the $Nd^{3+}$ wave functions and magnetic g-factors. From the observed splitting of the $Nd^{3+}$ ground state 8.8 cm$^{-1}$ at 5 K (in the easy-plane antiferromagnetic state of $NdFe_3(BO_3)_4$) and using the calculated value of g-factor in the ground state $g_\perp$ = 2.385, we obtained the following estimates for the values of the local effective magnetic field at the $Nd^{3+}$ site and the Nd – Fe exchange integral, respectively, $B_{loc}$=7.88 T and $|J_{FN}|$= 0.48 K. Further, we have analyzed the experimental data from literature on the magnetic susceptibility tensor of $NdFe_3(BO_3)_4$, considering the contributions from interacting neodymium and iron subsystems. For the latter, the strongest exchange interactions are those within Fe-O-Fe spiral chains. To account, at least partly, for these quasi-one-dimensional properties of the Fe subsystem, we took the exact form of the susceptibility of a dimer containing two nearest-neighbor iron ions in the chain (connected by the exchange interaction which was characterized by the parameter $J_{nn}$), and then used the mean-field approximation (with the parameter $J_{nnn}$ responsible for an effective inter-chain exchange interaction). The explicit form of the magnetic susceptibility was derived and the values of the exchange interaction parameters $J_{nn}$ = -6.25 K and $J_{nnn}$ = -1.92 K and the sign of the Nd-Fe exchange parameter $J_{FN}$ were obtained from fitting the calculated temperature dependence of the transversal susceptibility to the measured data. (We note that the signs of the considered exchange parameters correspond to the antiferromagnetic interactions between $Fe^{3+}$ ions and to the antiferromagnetic interaction between $Fe^{3+}$ ions and $Nd^{3+}$ ions in the ground state.) Then, the longitudinal susceptibility was calculated using the same parameters. A good agreement between theory and experimental data in the paramagnetic phase has been achieved. The main specific features of the temperature dependences of magnetic susceptibilities in the antiferromagnetic phase are also satisfactorily reproduced by calculations without introducing additional fitting parameters.

## ACKNOWLEDGEMENTS

This work was supported by the RFBR grants No 07-02-01185 and No 06-02-16088. BZM and ARZ acknowledge the support from the Ministry of Education and Science of Russian Federation (project RNP 2.1.1.7348).

**Appendix. Cluster susceptibilities in the staggered magnetic field**

Hamiltonian of a dimer in the staggered magnetic field $\mathbf{B}$ has the form

$$H^{ad} = -2J_{nn}\mathbf{S}_1\mathbf{S}_2 - g\mu_B(\mathbf{S}_1 - \mathbf{S}_2)\mathbf{B} - 2(J_{nn} + 2J_{nnn})<\mathbf{S}>(\mathbf{S}_1 - \mathbf{S}_2)$$

where average values of single ion spin moments ($S_1 = S_2 = S$) satisfy the condition $<S_1>$ = $-<S_2>$ = $-<S>$ ($J_{nn}<0$, $J_{nn}+2J_{nnn}<0$). The operator $\mathbf{S}_1-\mathbf{S}_2$ is zero within the manifolds corresponding to a fixed value of the total spin moment $S_t$. The non-zero average moment $<S_z>$ appears only due to mixing of states $|S_t,S_{tz}>$ and $|S_{t+1},S_{tz}>$. Using the analytical form of the matrix element $<S_1,S_2,S_t,S_{tz}|S_{1z}|S_1,S_2,S_{t+1},S_{tz}>$ [22], we obtain the effective antiferromagnetic susceptibility

$$\chi_{Fe}^{AF}(T) = \frac{(g\mu_B)^2 \lambda(T)}{2[(J_{nn} + 2J_{nnn})\lambda(T) - J_{nn}]}.$$

Here $\lambda(T) = [1 + 4S(S+1)]/3Z(T) - 1/3$, the statistical sum $Z(T)$ is given above by Eq. (10).

The effective single-ion susceptibility in the cluster containing three adjacent ions in the chain ($S_1 = S_2 = S_3 = S$) can be obtained similarly:

$$\chi_{Fe,t}^{AF} = \frac{(g\mu_B)^2 A}{1-B},$$

where

$$A = \sum_{S'=0}^{2S} \sum_{S_t=|S'-S|}^{S'+S} \frac{(2S_t+1)}{9Z_t kT} e^{J_{nn}S_t(S_t+1)/kT} \{S_t(S_t+1)(2\gamma_{S_tS'}-1)^2 - \frac{kT}{J_{nn}}[\left(\frac{(S'-S)(S'+S+1)}{S_t(S_t+1)}\right)^2 - 1]\},$$

$$B = \sum_{S'=0}^{2S} \sum_{S_t=|S'-S|}^{S'+S} \frac{(2S_t+1)}{9Z_t kT} e^{J_{nn}S_t(S_t+1)/kT} \times \{2S_t(S_t+1)(1-2\gamma_{S_tS'})[\gamma_{S_tS'}J_{nn} + 2(2\gamma_{S_tS'}-1)J_{nnn}] + \frac{kT}{J_{nn}}(J_{nn}+4J_{nnn})[\left(\frac{(S'-S)(S'+S+1)}{S_t(S_t+1)}\right)^2 - 1]\},$$

$$Z_t = \sum_{S'=0}^{2S} \sum_{S_t=|S'-S|}^{S'+S} (2S_t+1)e^{J_{nn}S_t(S_t+1)/kT},$$

$$\gamma_{S_tS'} = \frac{S_t(S_t+1) + S'(S'+1) - S(S+1)}{2S_t(S_t+1)}.$$

Table 1. Measured ($T$ = 50 K) and calculated crystal-field energies $E$ of Nd$^{3+}$ in NdFe$_3$(BO$_3$)$_4$, measured ($T$ = 5 K) exchange splittings $\Delta_{exp}$ of Kramers doublet states, and calculated $g$-factors.

| $^{2S+1}L_J$ | $E$ (cm$^{-1}$) Exper. | $E$ (cm$^{-1}$) Theory | $\Gamma_i$ | $\Delta_{exp}$ (cm$^{-1}$) | $g_\perp$ | $g_\parallel$ |
|---|---|---|---|---|---|---|
| $^4I_{9/2}$ | 0 | 0 | 4 | 8.8 | 2.385 | 1.376 |
|  | 65 | 66 | 56 |  | 0 | 3.947 |
| $3\Gamma_4 + 2\Gamma_{56}$ | 141 | 141 | 4 |  | 2.283 | 2.786 |
|  | 221 | 220 | 56 |  | 0 | 3.879 |
|  | 322 | 310 | 4 |  | 3.527 | 0.843 |
| $^4I_{11/2}$ | 1949 | 1945 | 4 |  | 3.708 | 3.508 |
|  | 1955 | 1954 | 56 |  | 0 | 8.03 |
| $4\Gamma_4 + 2\Gamma_{56}$ | 1979 | 1974 | 4 |  | 3.109 | 4.091 |
|  | 2053 | 2047 | 4 |  | 0.683 | 8.15 |
|  | 2087 | 2087 | 56 |  | 0 | 3.612 |
|  | 2131 | 2131 | 4 |  | 5.807 | 1.013 |
| $^4I_{13/2}$ | 3911 | 3912 | 4 | 1.4 | 4.810 | 5.174 |
|  | 3913 | 3917 | 56 | 8.8 | 0 | 9.265 |
| $5\Gamma_4 + 2\Gamma_{56}$ | 3933 | 3938 | 4 | 3 | 3.811 | 6.537 |
|  | 4012 | 4013 | 4 | 5 | 3.272 | 9.815 |
|  | 4038 | 4040 | 4 | 6.1 | 4.293 | 4.161 |
|  | 4083 | 4088 | 56 | 9 | 0 | 3.661 |
|  | 4129 | 4134 | 4 |  | 7.684 | 1.085 |
| $^4I_{15/2}$ | 5876 | 5877 | 56* | 1.7 | 0 | 10.81 |
|  | 5881 | 5887 | 4* | 3.5 | 7.140 | 4.505 |
| $5\Gamma_4 + 3\Gamma_{56}$ | 5949 | 5958 | 4 |  | 3.179 | 8.367 |
|  | 6045 | 6046 | 56* | 4.8 | 0 | 16.28 |
|  | 6056 | 6074 | 4* | 6.8 | 3.347 | 5.939 |
|  | — | 6132 | 4 ? |  | 7.077 | 2.482 |
|  | 6203 | 6197 | 56 |  | 0 | 3.669 |
|  | 6260 | 6261 | 4 |  | 9.309 | 1.496 |
| $^4F_{3/2}$ | 11349 | 11347 | 4 | 2.5 | 0.926 | 0.251 |
| $\Gamma_4 + \Gamma_{56}$ | 11430 | 11425 | 56 |  | 0 | 1.562 |
| $^4F_{5/2}$ | 12362 | 12355 | 4 |  | 3.158 | 0.598 |
| $2\Gamma_4 + \Gamma_{56}$ | 12441 | 12430 | 4* |  | 0.096 | 4.713 |
|  | 12454 | 12461 | 56* |  | 0 | 2.576 |
| $^2H_{9/2}$ | 12486 | 12503 | 4* | 2 | 3.995 | 0.982 |
|  | 12538 | 12542 | 4 |  | 1.989 | 4.633 |
| $3\Gamma_4 + 2\Gamma_{56}$ | 12560 | 12562 | 56 |  | 0 | 2.169 |
|  | 12605 | 12635 | 4* |  | 2.877 | 2.765 |
|  | 12697 | 12645 | 56* |  | 0 | 7.788 |
| $^4F_{7/2} + {}^4S_{3/2}$ | 13330 | 13346 | 4 | 1.9 | 4.895 | 1.296 |
|  | 13351 | 13356 | 56 |  | 0 | 3.524 |
| $3\Gamma_4 + \Gamma_{56}$ | 13420 | 13424 | 4 |  | 3.216 | 0.673 |
| $\Gamma_4 + \Gamma_{56}$ | 13472 | 13470 | 4* | 3.4 | 3.858 | 2.141 |
|  | 13483 | 13479 | 56* |  | 0 | 5.848 |
|  | — | 13495 | 4 ? |  | 3.239 | 1.884 |
| $^4F_{9/2}$ | 14592 | 14596 | 4 | 1 | 4.190 | 2.690 |
|  | 14623 | 14626 | 4 |  | 3.252 | 3.204 |
| $3\Gamma_4 + 2\Gamma_{56}$ | 14631 | 14631 | 56 |  | 0 | 4.157 |
|  | 14723 | 14711 | 4 |  | 5.208 | 3.139 |
|  | 14820 | 14809 | 56 |  | 0 | 10.80 |
| $^2H_{11/2}$ | — | 15835 | - | | 3.7268 | 5.798 |
| | 15840 | 15861 | 56 | | 0 | 7.931 |
| $4\Gamma_4 + 2\Gamma_{56}$ | — | 15862 | - | | 3.482 | 4.435 |
| | 15928 | 15946 | 56 | | 0 | 5.084 |
| | 15967 | 15957 | 4 | | 5.259 | 0.430 |
| | 15993 | 15991 | 4 | | 1.083 | 9.728 |
| $^4G_{5/2} + {}^2G_{7/2}$ | 16900 | 16912 | 4 | 8.4 | 0.043 | 0.065 |
| | 17045 | 17062 | 4 | 3.4 | 1.385 | 1.310 |
| $2\Gamma_4 + \Gamma_{56}$ | 17081 | 17088 | 56 | | 0 | 3.044 |
| $3\Gamma_4 + \Gamma_{56}$ | 17181 | 17179 | 4 | | 2.617 | 0.266 |
| | 17227 | 17218 | 4 | ~7 | 0.755 | 3.308 |
| | 17274 | 17248 | 4 | | 1.578 | 0.954 |
| | 17308 | 17291 | 56 | | 0 | 1.016 |
| $^4G_{7/2}$ | 18850 | 18822 | 4 | 2.2 | 3.369 | 1.784 |
| | 18892 | 18895 | 4 | | 2.098 | 3.026 |
| $3\Gamma_4 + \Gamma_{56}$ | 18936 | 18933 | 56 | | 0 | 2.924 |
| | 18993 | 18992 | 4 | 1.4 | 2.292 | 2.081 |
| $^4G_{9/2} + {}^2K_{13/2}$ | 19275 | 19275 | 56 | | 0 | 8.530 |
| | 19288 | 19272 | 4 | | 0.283 | 9.760 |
| $3\Gamma_4 + 2\Gamma_{56}$ | 19359 | 19360 | 4 | | 0.395 | 9.860 |
| $5\Gamma_4 + 2\Gamma_{56}$ | 19381 | 19396 | 56 | 1 | 0 | 5.086 |
| | 19388 | 19400 | 4 | | 4.017 | 1.087 |
| | 19418 | 19426 | 56* | | 0 | 7.030 |
| | 19443 | 19427 | 4* | | 1.429 | 2.529 |
| | — | 19482 | 4 ? | | 1.210 | 8.362 |
| | 19503 | 19495 | 4 | | 0.017 | 6.594 |
| | 19689 | 19629 | 4 | | 1.544 | 3.915 |
| | 19772 | 19717 | 56 | | 0 | 3.025 |
| | 19785 | 19753 | 4 | | 6.308 | 0.694 |
| $^2G_{9/2} + {}^4G_{11/2}$ | 20853 | 20840 | 4 ? | | 5.478 | 1.191 |
| $+ {}^2D_{3/2}$ | 20875 | 20872 | 56 ? | 2 | 0 | 5.675 |
| $+ {}^2K_{15/2}$ | 20948 | 20952 | 56 ? | 3 | 0 | 7.584 |
| | — | 20967 | 4 ? | | 2.757 | 3.680 |
| $3\Gamma_4 + 2\Gamma_{56}$ | — | 21034 | 4 ? | | 2.759 | 5.836 |
| $4\Gamma_4 + 2\Gamma_{56}$ | 21074 | 21072 | 4 | | 2.260 | 1.176 |
| $\Gamma_4 + \Gamma_{56}$ | 21120 | 21126 | 56 ? | | 0 | 3.564 |
| $5\Gamma_4 + 3\Gamma_{56}$ | 21201 | 21260 | 4 ? | | 0.049 | 12.45 |
| | 21232 | 21335 | 4 ? | | 6.726 | 0.052 |
| | 21317 | 21347 | 4 ? | | 0.377 | 12.52 |
| | 21347 | 21390 | 4 ? | | 0.878 | 10.31 |
| | — | 21403 | 56 ? | | 0 | 3.764 |
| | 21444 | 21443 | 56 ? | | 0 | 12.97 |
| | 21495 | 21458 | 56 ? | | 0 | 9.558 |
| | — | 21544 | 4 ? | | 0.122 | 5.428 |
| | — | 21580 | 4 ? | | 4.436 | 5.982 |
| | 21620 | 21643 | 4 ? | | 0.522 | 7.700 |
| | — | 21691 | 56 ? | | 0 | 10.5 |
| | — | 21695 | 4 ? | | 4.478 | 3.807 |
| | — | 21749 | 56 ? | | 0 | 3.784 |
| | 21777 | 21776 | 4 ? | | 8.132 | 0.975 |

*Experimental polarization properties do not allow assigning the IRREP unambiguously, but they do not contradict to the results of calculations
? IRREPs which have not been determined from measurements are marked by the "?" sign.

Table 2. Allowed transitions between the CF energy levels ($D_3$ point symmetry group).

|  | $\Gamma_4$ | $\Gamma_{56}$ |
|---|---|---|
| $\Gamma_4$ | $d_x, d_y, d_z; \mu_x, \mu_y, \mu_z$ <br> $\alpha, \sigma, \pi$ ED; $\alpha, \sigma, \pi$ MD | $d_x, d_y; \mu_x, \mu_y$ <br> $\alpha, \sigma$ ED; $\alpha, \pi$ MD |
| $\Gamma_{56}$ | $d_x, d_y; \mu_x, \mu_y$ <br> $\alpha, \sigma$ ED; $\alpha, \pi$ MD | $d_z, \mu_z$ <br> $\pi$ ED; $\sigma$ MD |